# Polarization Discrimination Imaging of objects hidden in turbid media: Detection of weak sinusoids through Stochastic Resonance

Samudra Dasgupta[1], Jithun Nair[1], Shauryadipta Sarkar[2], Ram Mohan Vasu[3] and Gargeshwari Venkatasubbiah Anand[4]

[1] *Department of Electronics and Electrical Communication Engineering, Indian Institute of Technology, Kharagpur 721 302, India.*

[2] *Department of Electrical Engineering, Indian Institute of Technology, Kharagpur 721 302, India.*

[3] *Department of Instrumentation, Indian Institute of Science, Bangalore 560 012, India.*

[4] *Department of Electrical Communication Engineering, Indian Institute of Science, Bangalore 560 012, India.*

*samudra@ieee.org, jithunnair@yahoo.co.in, shaurya_iit@yahoo.com, vasu@isu.iisc.ernet.in, anandgv@ece.iisc.ernet.in*

In Polarization Discrimination Imaging, the amplitude of a sinusoid from a rotating analyzer, representing residual polarized light and carrying information on the object, is detected with the help of a lock-in amplifier. When turbidity increases beyond a level, the lock-in amplifier fails to detect the weak sinusoidal component in the transmitted light. In this work we have employed the principle of Stochastic Resonance and used a 3-level quantizer to detect the amplitude of the sinusoids, which was not detectable with a lock-in amplifier. In using the three level quantizer we have employed three different approaches to extract the amplitude of the weak sinusoids: (a) using the probability of the quantized output to crossover a certain threshold in the quantizer (b) maximizing the likelihood function for the quantized detected intensity data and (c) arriving at an expression for the expected power in the detected output and comparing it with the experimentally measured power. We have proven these non-linear estimation methods by detecting the hidden object from experimental data from a polarization discrimination imaging system. When the turbidity increased to $L/l^{*} = 5.05$ ($l^{*}$ is the transport mean-free-path and $L$ is the thickness of the turbid medium) the data through analysis by the proposed methods revealed the presence of the object from the estimated amplitudes. This was not possible by using only the lock-in amplifier system. 

## 1. Introduction

There are two approaches to imaging of objects hidden in turbid media. One of them uses multiple scattered light employing an accurate model for propagation of such scattered light. The other uses the least scattered, ballistic or snake photons, which are assumed to propagate along straight line paths. The second method is limited to objects for which the ratio of thickness to transport mean-free-path is not large so that there are enough least scattered photons present in the transmitted light. For imaging using least-scattered photons, they need to be separated from the large background created by multiple-scattered light. Many methods are used to separate the least-scattered photons, such as time-gating[3] , heterodyne detection[4] , confocal detection[5] , coherent gating[6] , and polarization gating[2] . Whereas methods such as time-gating and confocal detection need expensive equipments, polarization gating leads to a relatively cheap and easily implementable alternative. The principle of polarization gating is that multiple-scattering destroys polarization and therefore the residual polarization in the transmitted light can be used as a discriminant to detect the least-scattered light.

The difficulty with polarization gating is that, when either the scattering coefficient, or the thickness of the medium increases, the polarization state of the output light becomes completely randomized and the residual polarized light, if there is any available, becomes too weak to be detected. To quantify the state of polarization in the transmitted light, a parameter, known as the scattering length[7] , defined as the ratio $L/l^*$, is used. Here $L$ is the thickness of the medium, and $l^*$ is the transport mean-free-path of photons in the medium. When $L/l^*$ is much less than a certain limit, known as diffusive length, ballistic or on-axis photons predominate in the transmitted light. When $L/l^*$ approaches the diffusive length, the least-scattered photons become very small. To facilitate detection of the residual polarized light, the output from the scattering object is sent through a rotating analyzer and the amplitude of the exiting sinusoidally varying intensity is detected with the help of a lock-in amplifier. When $L/l^*$ increases much beyond the diffusive length, even the lock-in amplifier will not be able to detect the weak sinusoid representing residual polarization. For thick biological objects for which $L/l^*$ is very large, the output light consists entirely of multiply scattered photons, and imaging based on least-scattered photons is not possible.

However, there is a range of $L/l^*$ values for which there still is a residual polarized component in the output, manifesting itself as a sinusoidal intensity too weak to be detected by the lock-in amplifier. Typical values of $L/l^*$ in this category are between 4.5 and 10.0, and beyond $L/l^* = 10$ one may presume there is no residual polarized component in the output. In this work we make use of a 3-level quantizer to extract the amplitude of the sinusoid in the output (representing the residual polarized component) making use of the principle of Stochastic Resonance (SR)[8] . The SR is a phenomenon in which the power from the noise spectrum is transferred to the signal when conditions are optimal for a noise assisted amplification of the signal. We have proposed and studied the relative performance of three different approaches to amplitude estimation making use of SR. With this SR-assisted estimation of the weak sinusoid we have shown, through simulation and experiment, that it is possible to detect objects hidden in scattering media even when the lock-in detection fails.

The rest of the paper is as follows: In Section 2 we briefly describe the experimental setup used for polarization discrimination imaging in a turbid medium. Here we also characterize the noise pedestal in the output intensity, coming from multiple scattering, for its probability density function. In Section 3 we have shown that the conventional ML estimator is crtically dependent on an exact knowledge of the frequency for estimation of the amplitudes of the weak sinusoids when the noise has a Rayleigh probability distribution. In practical situations often the frequency is not known with full accuracy. Hence there is a need for more robust estimators which are less sensitive to mild frequency variations. Section 4 describes the estimation methods used to detect the weak sinusoid buried in the background scattered intensity and also compares their relative performance. It is shown that the variation of the detected amplitude of the residual sinusoid provides a shadow of the object buried in the highly turbid medium when the alternative methods such as that using the lock-in amplifier or maximum likelihood estimator fail. Our conclusions are given in Section 5.

## 2. Experimental setup for polarization discrimination imaging

The set-up is shown schematically in Figure 1. The input illumination comes from a He-Ne laser whose spatial coherence is destroyed by sending it through a set of static (SD) and rotating (RD) diffusers. The RD has attached to it a plane polarizer. The light coming from RD is collimated by the lens which illuminates the object. The object is a glass rod immersed in a cuvette filled with a scattering medium consisting of polystyrene beads suspended in water. The exit plane of the cuvette is the image plane I where we want to measure the residual polarized light. The plane I is imaged by lenses L1 and L2 onto the output plane. The intermediate Fourier plane has a static analyzer. The output plane O is scanned by a photo-multiplier tube (PMT) which is interfaced to a computer. The PMT is scanned horizontally and for each position the time series data representing the residual sinusoid in noise is gathered. This is done for two scattering cross-sections of the turbid medium represented by $L/l^* = 2.14$ and 5.05. The data sets are analyzed using the 3-level quantizer as described in the next section.

*Noise Characterization*

Estimation of the intensity of polarized light requires characterization of the noise first. For this purpose we collect the data after the polarized light is passed through a highly turbid medium with $L/l^* = 5.05$ . The transmitted light almost completely loses its polarization and the power spectrum of the detected light is entirely white showing no sign of the presence of the sinusoudal intensity. For characterizing the noise, we collect a set of intensity vs time data along the image plane and plot their histograms. A typical histogram is shown in Figure 2 which clearly shows that the noise follows a Rayleigh distribution with pdf:

$$P(r) = \frac{r}{s^2}\exp(-\frac{r^2}{2s^2})u(r) \tag{1}$$

where $u(r)$ is the unit step function. We estimate the parameter 's' from each of the experimentally obtained histograms by fitting the curve of eqn. (1). The parameter 's' is the average of all such values extracted from the histograms, which is found to be 0.026.

## 3. Estimation of amplitude through Maximum Likelihood Estimator (MLE)

Estimation of parameters of a weak sinusoid in noise has attracted the attention of many workers in the past. The existing methods can be classified under two heads: (i) techniques which employ a Bayesian approach and (ii) those which employ the classical approach wherein the unknown parameters are considered deterministic constants. In the Bayesian approach, the parameters to be estimated are considered as random variables whose probability distributions are assumed *a priori* known.

The Bayesian approach is unsuitable for our application since we do not have any knowledge about the pdf of the amplitude of the noisy sinusoid. Therefore we turned our attention to such approaches as those using the Cramer-Rao Lower Bound (CRLB), Rao-Blackwell-Lehmann-Scheffe method, Best Linear Unbiased Estimator (BLUE), Maximum Likelihood Estimator (MLE), Least Squares Estimator (LSE) and Method of moments which come under the classical methods. Of these we have dropped both the LSE and the method of moments because their optimality with respect to variance cannot be analysed in general. Further it can be proven that there is no amplitude estimator for a weak sinusoid in Rayleigh noise that is unbiased and attains the CRLB. The Rao-Blackwell-Lehmann-Scheffe estimator will produce the Minimum Variance Unbiased Estimator if and only if a complete and sufficient statistic is available. (A statistic is complete if there is only one function of the statistic that is unbiased.) The BLUE has the restriction that the estimator be linear. It selects that linear estimator which ensures minimum variance which may not be the best estimator. Because of these reasons we have rejected all these classical approaches except the MLE. In what follows we describe the application of MLE for estimation of amplitude from the measured noisy intensity data.

### *3.A. Amplitude estimation through MLE*

The noisy data are simulated by adding Rayleigh distributed noise to weak sinusoids . The signal to noise ratio (SNR) is kept low in the range varying from $-23$dB to $-43$dB. We have then tested the ML estimator by comparing the estimated amplitude with the ones used in simulating the data. The details are as follows:

The simulated noisy input signal x(n) is given by:

$$x_n = \sigma\{A\cos(2\pi f_0 n + \phi) + w_n\}, \qquad n = 0, 1, \ldots N-1 \tag{2}$$

where $\sigma^2$ is the noise variance, $w_n$ is a unit-variance random variable with Rayleigh distribution, A is the normalized amplitude, $f_0$ is the frequency of operation and $\phi$ is a random phase. The probability distribution function of $w_n$ is given by:

$$f(w_n) = a^2 w_n \exp\left(-\frac{a^2 w_n^2}{2}\right) \tag{3}$$

where $a = \sqrt{2 - \frac{\pi}{2}}$. For a particular $A$ and $\phi$, it follows that the probability distribution of $x_n$ is given by :

$$f(x;A,\phi) = \left(\frac{a}{\sigma}\right)^{2N} \prod_{n=0}^{N} [x_n - A\cos(2\pi f_0 n + \phi)] \exp\left[-\frac{a^2}{2\sigma^2} \sum_{n=0}^{N} \{x_n - A\cos(2\pi f_0 n + \phi)\}^2\right] \quad (4)$$

Assuming that the number of data points is large and following the standard procedure for ML estimation, we have the following implicit formulae for the estimate of amplitude, $\hat{A}$:

$$\hat{\alpha}_c = \frac{2}{N} \sum_{n=0}^{N-1} \left( x_n - \frac{\sigma^2/a^2}{x_n - \hat{\alpha}_c \cos(2\pi f_0 n) - \hat{\alpha}_s \sin(2\pi f_0 n)} \right) \cos(2\pi f_0 n) \quad (5)$$

$$\hat{\alpha}_s = \frac{2}{N} \sum_{n=0}^{N-1} \left( x_n - \frac{\sigma^2/a^2}{x_n - \hat{\alpha}_c \cos(2\pi f_0 n) - \hat{\alpha}_s \sin(2\pi f_0 n)} \right) \sin(2\pi f_0 n) \quad (6)$$

and,

$$\hat{A} = \sqrt{\hat{\alpha}_c^2 + \hat{\alpha}_s^2} \quad (7)$$

We use the following initial values for $\hat{\alpha}_s$ and $\hat{\alpha}_c$:

$$\hat{\alpha_{c,0}} = \frac{2}{N} \sum_{n=0}^{N-1} x_n \cos(2\pi f_0 n) \quad (8)$$

$$\hat{\alpha_{s,0}} = \frac{2}{N} \sum_{n=0}^{N-1} x_n \sin(2\pi f_0 n) \quad (9)$$

We have repeated the estimation of amplitudes for a set of simulated data where the SNR is varied from $-23$dB to $-43$dB. Figure 3 shows the error in the estimated amplitudes. It is seen that the MLE performs extremely well with the error showing forth only a small increase for very low SNRs. Here we have assumed full knowledge of the frequency $f_0$. However if there is an uncertainty in the frequency even by a small amount, say by 0.05%, the error of estimation of amplitude increases as large as 87%. In practical situations such as the one considered in this work, where there is bound to be frequency uncertainties, this method is not applicable. For example, when this method is applied to our experimental data (which is basically the transmitted intensity from a glass rod in turbid medium) the distribution of estimated amplitudes does not reveal the presence of the object (Figure 4). Therefore there is a need for a more robust amplitude estimator to deal with situations where frequency is not known very accurately. In the next section we describe three more robust amplitude estimators based on SR employing three level quantizers.

## 4. Amplitude estimation employing Stochastic Resonance in a three level quantizer

In the context of estimation of weak periodic signals the nonlinear phenomenon of Stochastic Resonance can be introduced as follows: When a periodic signal superposed with noise is input to a nonlinear system, a cooperative transfer of power from noise to signal takes place when the noise is adjusted to a certain optimal level. Considering that favourable conditions have to be tailored to promote this transfer of power the name resonance is justified. Experimental evidence of SR has been reported in various systems [9–14]. To bring in resonance and the noise-assisted SNR improvement, apart from the level of noise in the input we may have also another parameter of the nonlinearity we can tailor. And this parameter, present in the three level tristable nonlinearity (which is not available in all nonlinearities) is the threshold. If we vary the threshold of quantization and plot the output SNR gain achieved through quantization, then we observe a non-monotonic variation of the SNR gain which gets maximized at a particular threshold where the SNR gain is greater than unity. The same SNR gain peaking can also be observed if we redid the above by keeping the threshold constant and increasing the input noise variance. In our case the input noise variance is already above the optimal point for achieving resonance and hence we must adjust the threshold to ensure an optimal enhancement of SNR in the output. For the case when the input SNR is low i.e. $\frac{A}{\sigma} \leq 0.1$, which is true in our case, it is possible to derive a theoretical expression for this optimal threshold for resonance.

The input signal $x_n$, as before, is given by:

$$x_n = \sigma\{A\cos(2\pi f_0 n) + w_n\}$$

where $w_n$ is a unit variance Rayleigh noise sequence with mean $m = \sqrt{\frac{\pi}{4-\pi}}$. Our 3-level uniform quantizer is defined as:

$$y_n = \begin{cases} 1 & if \quad x_n > \sigma(\gamma + m) \\ 0 & if \quad \sigma(-\gamma + m) < x_n < \sigma(\gamma + m) \\ -1 & if \quad x_n < \sigma(-\gamma + m) \end{cases} \tag{10}$$

$\gamma$ is the normalized threshold (normalized with respect to the noise variance $\sigma$). The expected output sequence $E[y_n]$ is given by:

$$\begin{aligned} E[y_n] &= P(y_n = 1) - P(y_n = -1) \\ &= P(w_n > \gamma + m - A\cos(2\pi f_0 n)) - P(w_n < -\gamma + m - A\cos(2\pi f_0 n)) \\ &= 1 - F(\gamma + m - A\cos(2\pi f_0 n)) - F(-\gamma + m - A\cos(2\pi f_0 n)) \end{aligned} \tag{11}$$

where $P(.)$ denotes the probability. Let us define

$$F_+ := F(\gamma + m) \tag{12}$$

$$
\begin{aligned}
F_- &:= F(-\gamma + m) & (13)\\
f_+ &:= f(\gamma + m) & (14)\\
f_- &:= f(-\gamma + m) & (15)
\end{aligned}
$$

where F(.) denotes the cumulative distribution function and f(.) denotes the probability distribution function. If $A << 1$,

$$E[y_n] \approx 1 - F_+ + A\cos(2\pi f_0 n) f_+ - F_- + A\cos(2\pi f_0 n) f_- \quad (16)$$

The variance of the sequence $y_n$ is given by:

$$
\begin{aligned}
\sigma_y^2 &= E[y_n^2] - \{E[y_n]\}^2\\
&\approx 1 - F_+ + A\cos(2\pi f_0 n) f_+ + F_- - A\cos(2\pi f_0 n) f_- -\\
&\quad \{(1 - F_+ - F_-)^2 + 2A\cos(2\pi f_0 n)(f_+ + f_-)(1 - F_+ - F_-)\} \quad (17)
\end{aligned}
$$

The time-average of the variance of $y_n$ is given by:

$$< \sigma_y^2 >= (1 - F_+ + F_-) - (1 - F_+ - F_-)^2 \quad (18)$$

It can be proved[18] that the output SNR ($\mu$) is given by the ratio of the sum of the signal powers at the frequencies $f_0, 2f_0, \ldots$ to the time-averaged variance of the output signal. Thus we arrive at the expression for $\mu$ as,

$$\mu = \frac{A^2(f_+ + f_-)^2}{2[(1 - F_+ + F_-) - (1 - F_+ - F_-)^2]} \quad (19)$$

Figure 5 is a plot of the output SNR $\mu$ vs threshold for Rayleigh noise. It exhibits a distinct peak. We choose the threshold corresponding to this maxima as the optimal threshold. To find the optimal threhold, we set $\frac{d\mu}{d\gamma} = 0$ so that

$$\frac{f_+ + f_-}{2(f'_+ + f'_-)} = \frac{F_+ + 3F_- - (F_+ + F_-)^2}{f_+ - 3f_- - 2(F_+ + F_-)(f_+ - f_-)} \quad (20)$$

This equation when solved numerically gives $\gamma_{opt} = 1.064$. From eqn. (10) the optimal threshold is $1.064\sigma$. In order to check whether the optimal threshold we have obtained above is the correct one corresponding to the maximum output SNR we have done the following numerical experiment: We have repeatedly estimated the optimal threshold in a three level quantizer for a peak in the output SNR. The SNR is evaluated by dividing the power in the fundamental and higher harmonics by the power every where else in the output power spectrum. The estimated optimal threshold is plotted in Figure 7 which averages around 1.046 at $\sigma = 1$ which is close to the value obtained from eqn. (20) i.e. 1.064.

We have also verified whether the assumption of very low SNR used in the above estimation of optimal threshold is valid in respect of the two experimental data sets we have gathered from the object in the turbid madium at $L/l^* = 5.05$ and $L/l^* = 2.14$. This we did by checking the power at $f_0 = 16.67Hz$, which is twice the rotation frequency of the polarizer, for the power spectra of the collected data. Whereas in the case of the experiment with $L/l^* = 2.14$ there was a peak at $f_0 = 16.67Hz$, this was absent for the case $L/l^* = 5.05$. Therefore we conclude that the low SNR assumption and the calculated optimal threshold are valid only for the case of experiment where $L/l^* = 5.05$ and not for the other. For $L/l^* = 2.14$ we must ascertain the optimal threshold by repeating the numerical experiments reported earlier with this experimental data. At the low turbidity level of $L/l^* = 2.14$, the glass rod can be detected using the amplitudes read out by the lock-in amplifier. This result is shown in Figure 6.

In the following we introduce three methods to estimate the amplitude of the weak sinusoid in the data analyzed by the three level quantizer.

### 4..1. Amplitude estimation through determining the crossover probability

Let $N_+$ and $N_-$ denote the number of samples that cross the upper and lower thresholds of quantization respectively. Also $N_0$ is the number of samples that do not cross either of the thresholds. Thus $N = N_0 + N_+ + N_-$ represents the total number of samples in the data. We shall find analytically the probability that the signal crosses the lower threshold of quantization i.e. $E(\frac{N_-}{N})$. If we use y to denote the output of the quantizer and T to denote the time period of the sinusoid, we can write,

$$\begin{aligned} E\left(\frac{N_-}{N}\right) &= P(y=-1) \\ &= \frac{1}{N}\sum_{n=0}^{N-1} F(-\gamma + m - A\cos(2\pi f_0 n)) \\ &= \left\langle 1 - \exp\left(-\frac{(-\gamma + m - A\cos(2\pi f_0 n))^2}{2s^2}\right)\right\rangle \\ &= 1 - \frac{1}{T}\int_0^T \exp\left(-\frac{(-\gamma + m - A\cos(\omega t))^2}{2s^2}\right) dt \qquad (21) \end{aligned}$$

where ¡.¿ denotes average over time. Now if we take a sufficiently long time series, then $\frac{N_-}{N}$ gives us the probability of lower crossover[1]. Equating it to $E\left(\frac{N_-}{N}\right)$ we can solve the equation above numerically to get the amplitude. Figure 8 shows the performance of this method to estimate the amplitudes. The error in amplitude estimation is very small if an exact knowledge of frequency is assumed. When there is an uncertainty in frequency of say 0.05%, the error in amplitude estimation increases but stays limited to 9%. Thus the method is relatively robust with respect to frequency variations.

[1] We could as well have worked with upper crossover probability or probability of no crossover. The analysis would follow a similar approach.

When the SNR is low, a closed form solution for the amplitude can be obtained from the expression for the probability for crossover:

$$\begin{aligned} E\left(\frac{N_-}{N}\right) &= < P(y_n = -1) > \\ &= \langle F(-\gamma + m - A\cos(2\pi f_0 n))\rangle \\ &\approx < F_- - A\cos(2\pi f_0 n) f_- + \frac{1}{2}A^2\cos^2(2\pi f_0 n) f'_- > \\ &\approx F_- + \frac{1}{4}A^2 f'_- \end{aligned} \tag{22}$$

where $F_-$ and $F_-$ have already been defined before and $f'_-$ represents differentiation of $f_-$ with respect to $\gamma$. This gives us

$$A = \sqrt{\frac{4[\frac{N_-}{N} - F_-]}{f'_-}} \tag{23}$$

The amplitude of the sinusoid is $A\sigma$ from eqn.(2). Figures 9 and 10 show plots of the estimated amplitudes vs position of the PMT for the cases when $L/l^* = 2.14$ and $L/l^* = 5.05$ respectively. They clearly reveal the presence of the glass rod. The higher amplitude at the middle of the rod is owing to the focusing action of the glass rod. The sudden drop in the amplitude near the edges of the rod help us locate it easily. These results can be compared to those obtained from measuring amplitudes through the lock-in amplifier (Figure 6). The results for the $L/l^* = 2.14$ case match. For the other case i.e. $L/l^* = 5.05$, whereas the lock-in amplifier failed, the present method succeeded in detecting the presence of the glass rod.

### 4..2. Amplitude estimation through application of the Maximum Likelihood Estimator to the quantized data

In this method we maximize the likelihood of obtaining the experimentally observed data after quantization. The likelihood function $L$ is defined as:

$$L = \prod_{n=1}^{N} P(y_n = \bar{y_n}) \tag{24}$$

where $\bar{y_n}$ is the experimentally obtained output sequence. As before,

$$P(y_n = 1) \quad = \quad 1 - F(\gamma + m - A\cos(2\pi f_0 n))$$

$$
\begin{aligned}
P(y_n = -1) &= F(-\gamma + m - A\cos(2\pi f_0 n)) \\
P(y_n = 0) &= 1 - P(y_n = 1) - P(y_n = -1)
\end{aligned}
$$

The particular value of $A$ which maximizes $L$ is the desired solution, which is found iteratively. Figure 11 shows the performance of ML estimation for quantized data. When the frequency is exactly known the performance of the ML estimator is excellent. With an uncertainty of 0.05% in the knowledge of frequency the performance drops and the error in amplitude estimation climbs up to 35%.

When the SNR is low, just as we have done in the method using the crossover probability, one can arrive at a closed form expression for the amplitude.
Thus,

$$
\begin{aligned}
< P(y_n = 1) > &= < 1 - F(\gamma + m - A\cos(2\pi f_0 n)) > \\
&\approx < 1 - \{F_+ - A\cos(2\pi f_0 n) f_+ + \frac{1}{2} A^2 \cos^2(2\pi f_0 n) f'_+ > \\
&\approx 1 - F_+ - \frac{1}{4} A^2 f'_+ \qquad (25)
\end{aligned}
$$

Similarly,

$$
\begin{aligned}
< P(y_n = -1) > &\approx F_- + \frac{1}{4} A^2 f'_- \qquad (26) \\
< P(y_n = 0) > &= 1 - < P(y_n = 1) > - < P(y_n = -1) > \\
&\approx (F_+ - F_-) + \frac{1}{4} A^2 (f'_+ - f'_-) \qquad (27)
\end{aligned}
$$

Differentiating the log likelihood function with respect to A and setting it to zero we get,

$$
\frac{d}{dA}\left\{ N_+ \log\left[1 - F_+ - \frac{1}{4}A^2 f'_+\right] + N_- \log\left[F_- + \frac{1}{4}A^2 f'_-\right] + N_0 \log\left[F_+ - F_- + \frac{1}{4}A^2(f'_+ - f'_-)\right]\right\} = 0 \qquad (28)
$$

This gives,

$$
-\frac{N_+ f'_+}{1 - F_+ - \frac{1}{4}A^2 f'_+} + \frac{N_- f'_-}{F_- + \frac{1}{4}A^2 f'_-} + \frac{N_0 (f'_+ - f'_-)}{F_+ - F_- + \frac{1}{4}A^2 (f'_+ - f'_-)} = 0 \qquad (29)
$$

This can be solved for A which gives us:

$$
A = \sqrt{\frac{-b - \sqrt{b^2 - 4ac}}{2a}} \qquad (30)
$$

where

$$
\begin{aligned}
a &= -\frac{Nf'_+f'_-(f'_+ - f'_-)}{16} \\
b &= \frac{1}{4}[-f'_+f'_-(F_+ - F_-)(m+n) + (f'_+ - f'_-)\{-mf'_+F_- + nf'_-(1-F_+) + \\
&\quad (N-n-m)(f'_-(1-F_+) - F_-f'_+)\}] \\
c &= (F_+ - F_-)[-mf'_+F_- + nf'_-(1-F_+)] + F_-(N-n-m)(f'_+ - f'_-)(1-F_+)
\end{aligned}
$$

The amplitude of the sinusoid is $A\sigma$ from eqn.(2). The Figure 12 shows the plot of the detected amplitudes vs PMT position revealing the glass rod when $L/l^* = 2.14$. Figure 13 is a similar plot revealing the glass rod when $L/l^* = 5.05$. We would like to point out that the iterative method employing the MLE failed to detect the glass rod when $L/l^* = 5.05$ whereas the closed form expression for amplitude obtained from log likelihood function succeeded.

These results can be compared to those obtained from measuring amplitudes through the lock-in amplifier (Figure 6). The results for the $L/l^* = 2.14$ case match. For the other case i.e. $L/l^* = 5.05$, whereas the lock-in amplifier failed, the present method employing the closed form solution for amplitude succeeded in detecting the presence of the glass rod.

### 4..3. Amplitude estimation through arriving at the expected power output from the quantizer

In this method we derive an analytical expression for the expected output power from the quantizer and equate it to that obtained by sending the experimental data through the quantizer. The expected power $E(P)$ is give by,

$$
\begin{aligned}
E(P) &= P(y_n = 1) + P(y_n = -1) \\
&= \exp\left(-\frac{(\gamma + m - Acos(2\pi f_0 n))^2}{2s^2}\right) + \left[1 - \exp\left(-\frac{(-\gamma + m - Acos(2\pi f_0 n))^2}{2s^2}\right)\right]
\end{aligned}
\tag{31}
$$

This is equated to the power calculated from the experimental data. The equation so obtained is iteratively solved for the desired amplitude. Figure 14 shows the performance of this method. As seen from the figure this method is the least sensitive to uncertainties in the frequency of the sinusoid. For example, when there is a change in frequency by 0.05% the estimation error goes up to only 2.5%.

Figure 15 and 16 show the plots of estimated amplitude vs PMT position for the two cases when $L/l^* = 2.14$ and $L/l^* = 5.05$ respectively. The shadow of the glass rod is clearly seen in both the cases.

Once again these results can be compared to those obtained from measuring amplitudes through the lock-in amplifier (Figure 6). The results for the $L/l^* = 2.14$ case match. For the other case i.e. $L/l^* = 5.05$, whereas the lock-in amplifier failed, the present method employing the power of the quantized output succeeded in detecting the presence of the glass rod.

## 5. Conclusion

Imaging in turbid media using polarization descrimination requires estimating the amplitudes of weak sinusoids representing residual polarization which is buried in noise of scattered light. This detection, usually done through lock-in amplifier, fails when the turbidity level indicated by $L/l^*$ values go beyond 4.5. In this work we have first shown that, one of the existing methods using a linear estimator and a likelihood function is critically dependent on the exact prior knowledge of frequency for its success when the noise has a Rayleigh distribution. We have devised three methods of estimation of amplitudes of weak sinusoids buried in noise, namely, that which uses the crossover probability, the maximum likelihood estimator for quantized data and the one using the expected power in the detected quantized output. The methods employ Stochastic Resonance to extract the amplitude information and they are all successful in detecting the presence of an object in a highly turbid medium. In addition, all are shown to be more robust in performance with respect to frequency uncertainties when compared to the existing linear estimators. Of the three methods, the one using the expression for expected output power is the least sensitive to frequency variations and hence most useful in practical situations.

### Acknowledgement

We thank Mr. Padmaram, Mr. Madhav, Mrs. Usha and Mr. Ganesh Chandan, from the Optical Tomography Laboratory, Department of Instrumentation, Indian Institute of Science, Bangalore, for their valuable help.

**List of Figure Captions**

Fig. 1. The experimental setup used for Polarization Discrimination Imaging. L is the Laser, S.D. and R.D. are the static and rotating diffusers, CL is the collimating lens, L1 and L2 are the imaging lenses.

Fig. 2. The normalized histogram of the noise shows that it follows a Rayleigh distribution. The 's' parameter of the Rayleigh pdf is estimated from a best-fit curve to the above histogram.

Fig. 3. A plot of the percentage error in estimated amplitude vs SNR in the input in the case of the linear estimator employing MLE. The dotted line represents the case when the frequency of the sinusoid is known exactly. The solid line is for the case when frequency is known only within an error of 0.05%.

Fig. 4. The plot of the estimated amplitude vs the position of the PMT in the case of amplitudes estimated by the linear method employing MLE. The $L/l^*$ of the turbid medium is 2.14. The pattern does not show forth the presence of the glass rod.

Fig. 5. Plot of equation 19 for output SNR $\mu$ vs threshold $\gamma$. The maximum SNR occurs around $\gamma = 1.064$.

Fig. 6. The plot of the estimated amplitude vs position of the PMT along the cuvette in the case of lock-in detection. The turbidity level is given by $L/l^* = 2.14$. The presence of the glass rod is clearly visible.

Fig. 7. Plot of the optimal threshold for the quantizer obtained for one hundred samples of the simulated data. The SNR in the data is kept at -23dB and the noise variance is kept at unity. The average value of the optimal threshold obtained is 1.046 which can be compared to the value obtained from eqn. (20) which is 1.064.

Fig. 8. The plot of the percentage error in the estimated amplitude vs SNR in the input from the 3 level quantizer employing the crossover probability method. The dotted line represents the case when the frequency of the sinusoid is known exactly. The solid line is for the case when frequency is known only within an error of 0.05%.

Fig. 9. The plot of the estimated amplitude vs position of the PMT along the cuvette in the case of amplitude estimated from the 3 level quantizer employing cross-over probability method. The turbidity level is given by $L/l^* = 2.14$. The presence of the glass rod is clearly visible.

Fig. 10. The same as Figure 9 except $L/l^* = 5.05$. The presence of the glass rod is still clearly visible.

Fig. 11. The plot of the percentage error in the estimated amplitude vs SNR in the input for the method employing the MLE for quantized data. The dotted line represents the case when the frequency of the sinusoid is known exactly. The solid line is for the case when frequency is known only within an error of 0.05%.

Fig. 12. The plot of the estimated amplitude vs position of the PMT along the cuvette in the case of amplitude estimated from the MLE for quantized data. The turbidity level is given by $L/l^* = 2.14$. The presence of the glass rod is clearly visible.

Fig. 13. The same as Figure 12 except $L/l^* = 5.05$. The presence of the glass rod is still clearly visible.

Fig. 14. The plot of the percentage error in the estimated amplitude vs SNR in the input employing the expected power of the quantized output. The dotted line represents the case when the frequency of the sinusoid is known exactly. The solid line is for the case when frequency is known only within an error of 0.05%.

Fig. 15. The plot of the estimated amplitude vs position of the PMT along the cuvette in the case of amplitude estimated from the 3 level quantizer employing the expected power of the quantized output method . The turbidity level is given by $L/l^* = 2.14$. The presence of the glass rod is clearly visible.

Fig. 16. The same as Figure 15 except $L/l^* = 5.05$. The presence of the glass rod is still clearly visible.

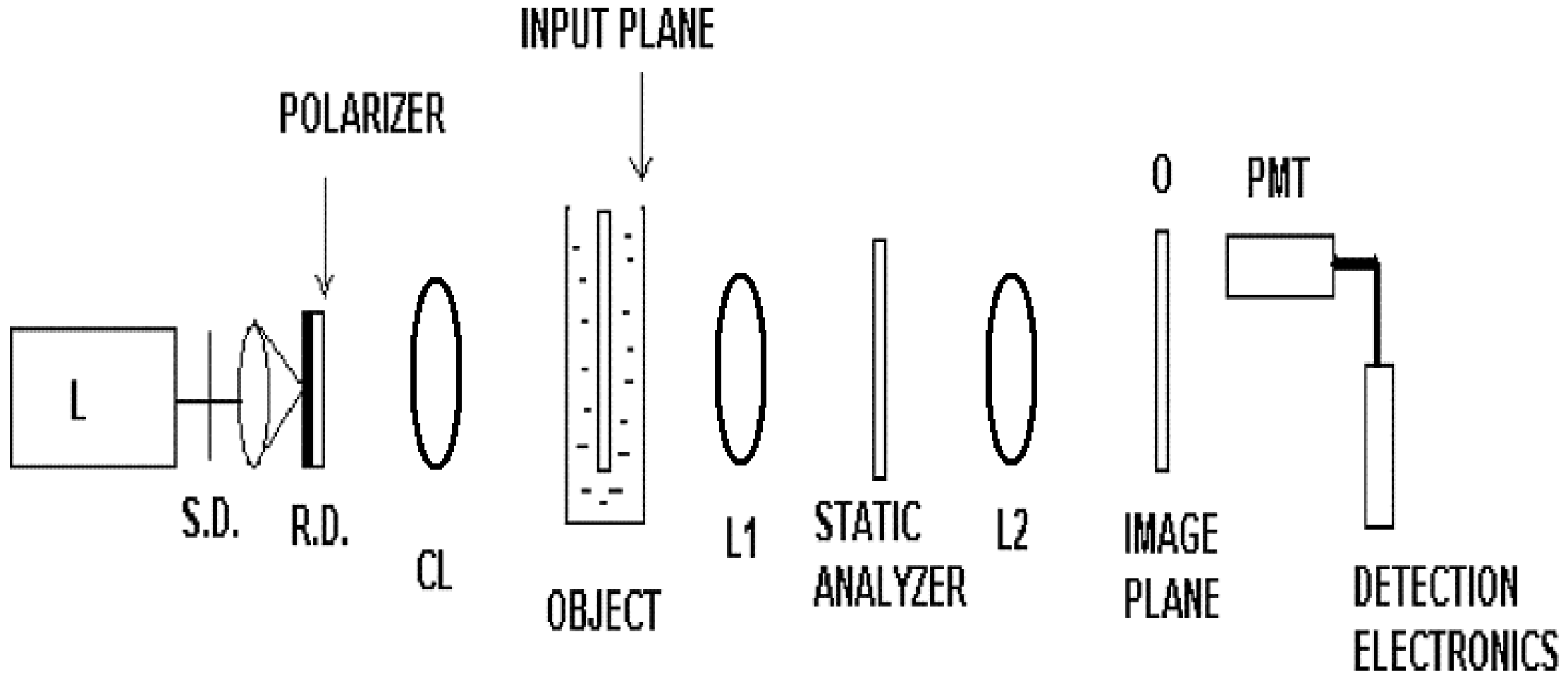

Fig. 1. The experimental setup used for Polarization Discrimination Imaging. L is the Laser, S.D. and R.D. are the static and rotating diffusers, CL is the collimating lens, L1 and L2 are the imaging lenses.

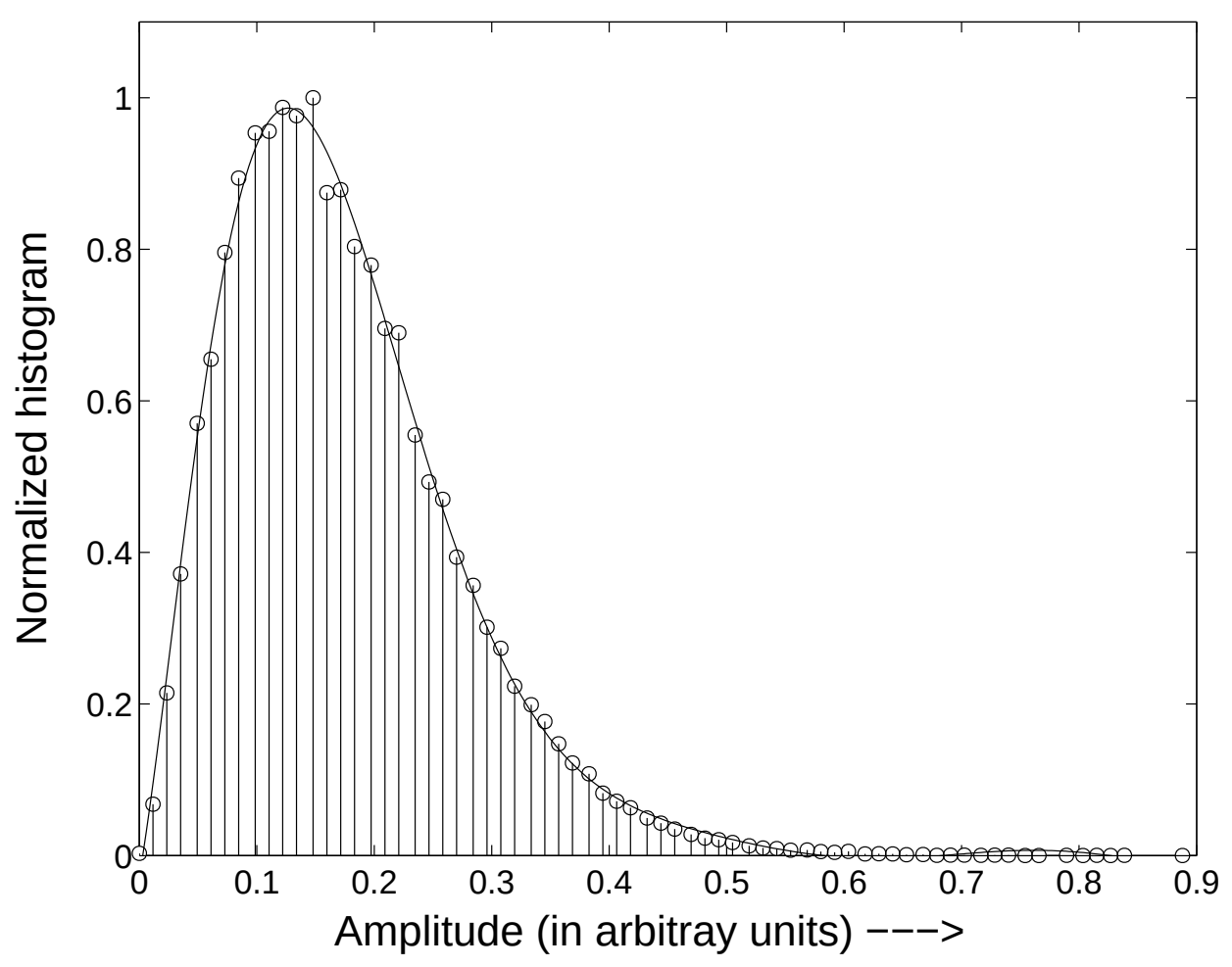

Fig. 2. The normalized histogram of the noise shows that it follows a Rayleigh distribution. The 's' parameter of the Rayleigh pdf is estimated from a best-fit curve to the above histogram.

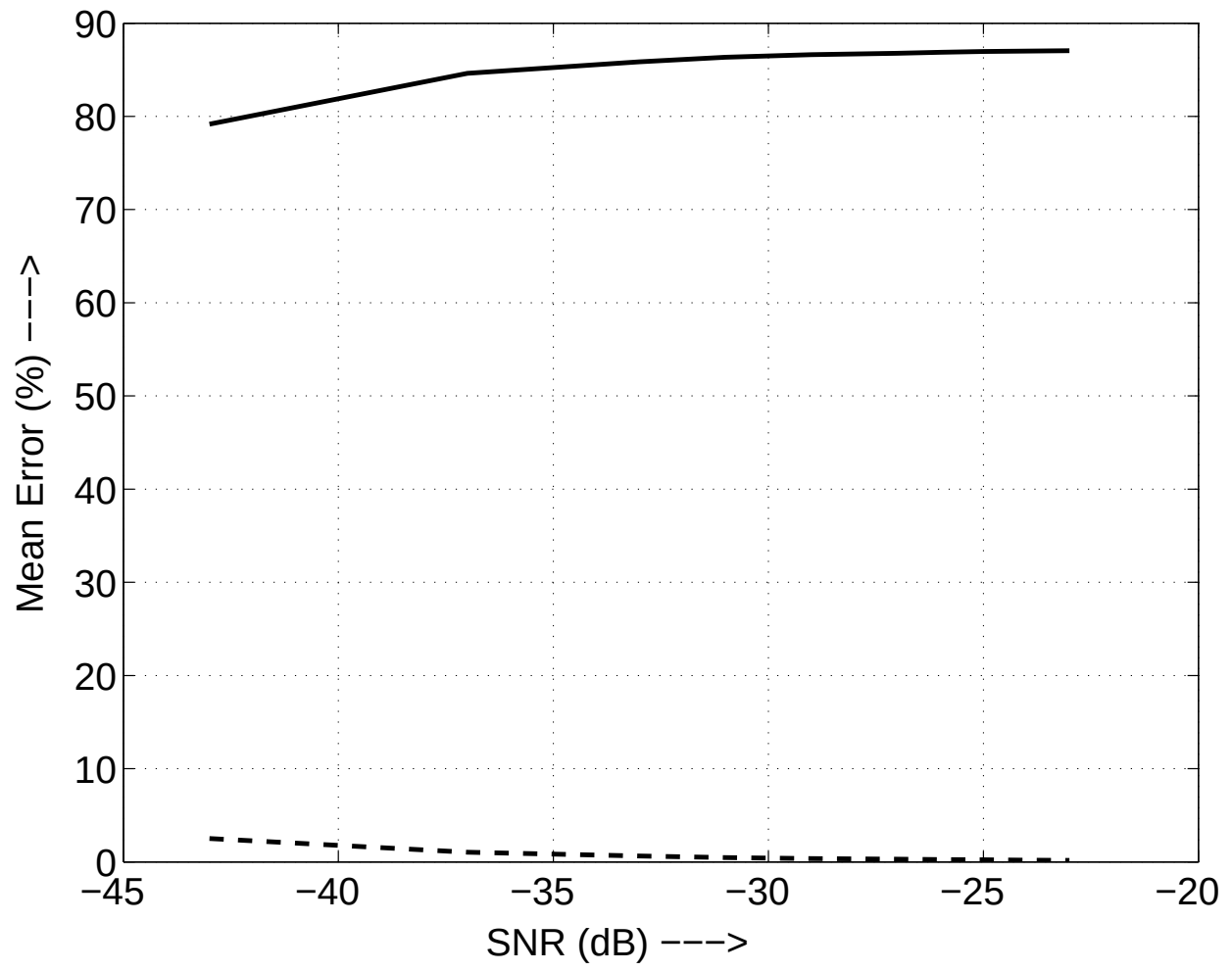

Fig. 3. A plot of the percentage error in estimated amplitude vs SNR in the input in the case of the linear estimator employing MLE. The dotted line represents the case when the frequency of the sinusoid is known exactly. The solid line is for the case when frequency is known only within an error of 0.05%.

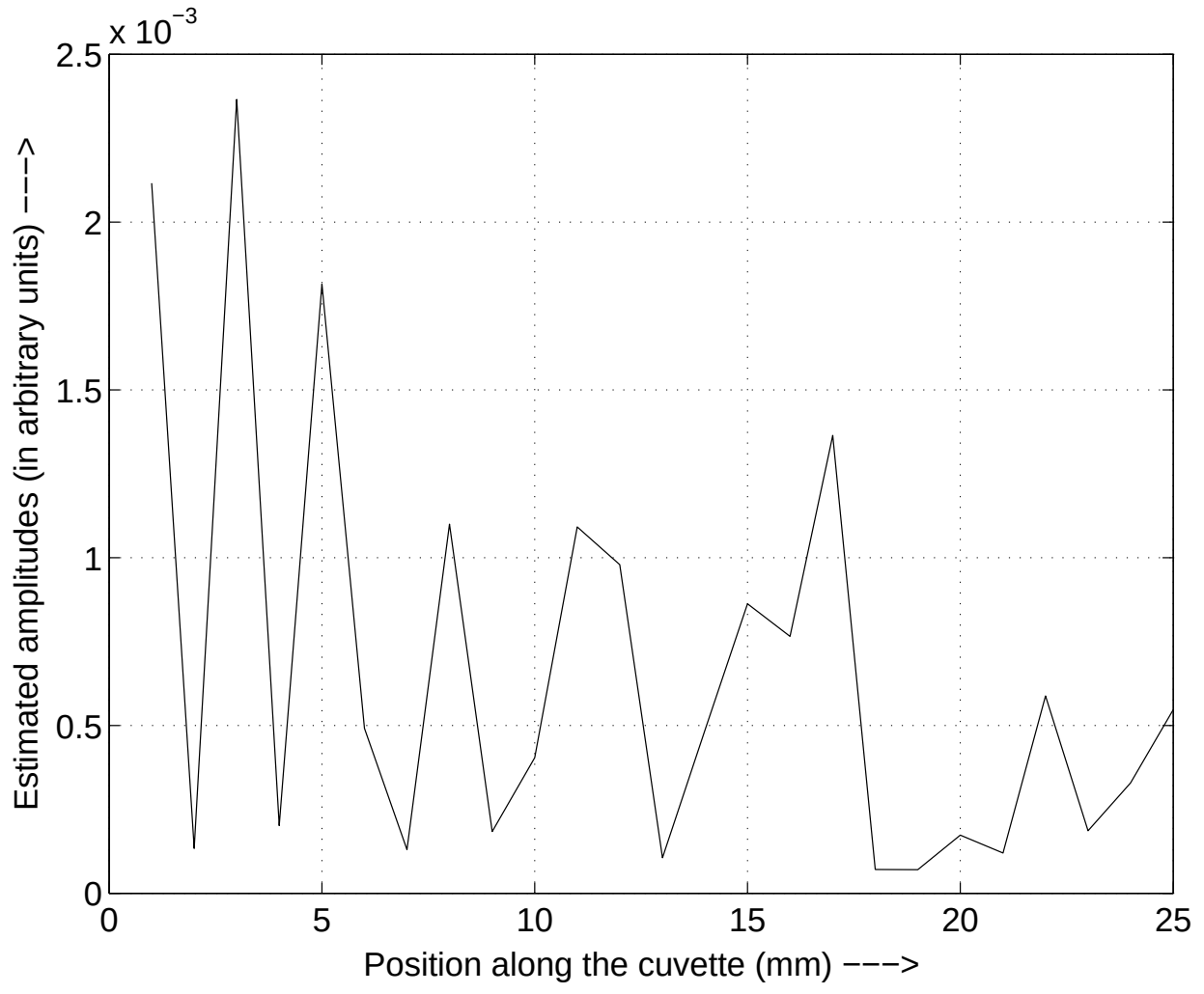

Fig. 4. The plot of the estimated amplitude vs the position of the PMT in the case of amplitudes estimated by the linear method employing MLE. The $L/l^*$ of the turbid medium is 2.14. The pattern does not show forth the presence of the glass rod.

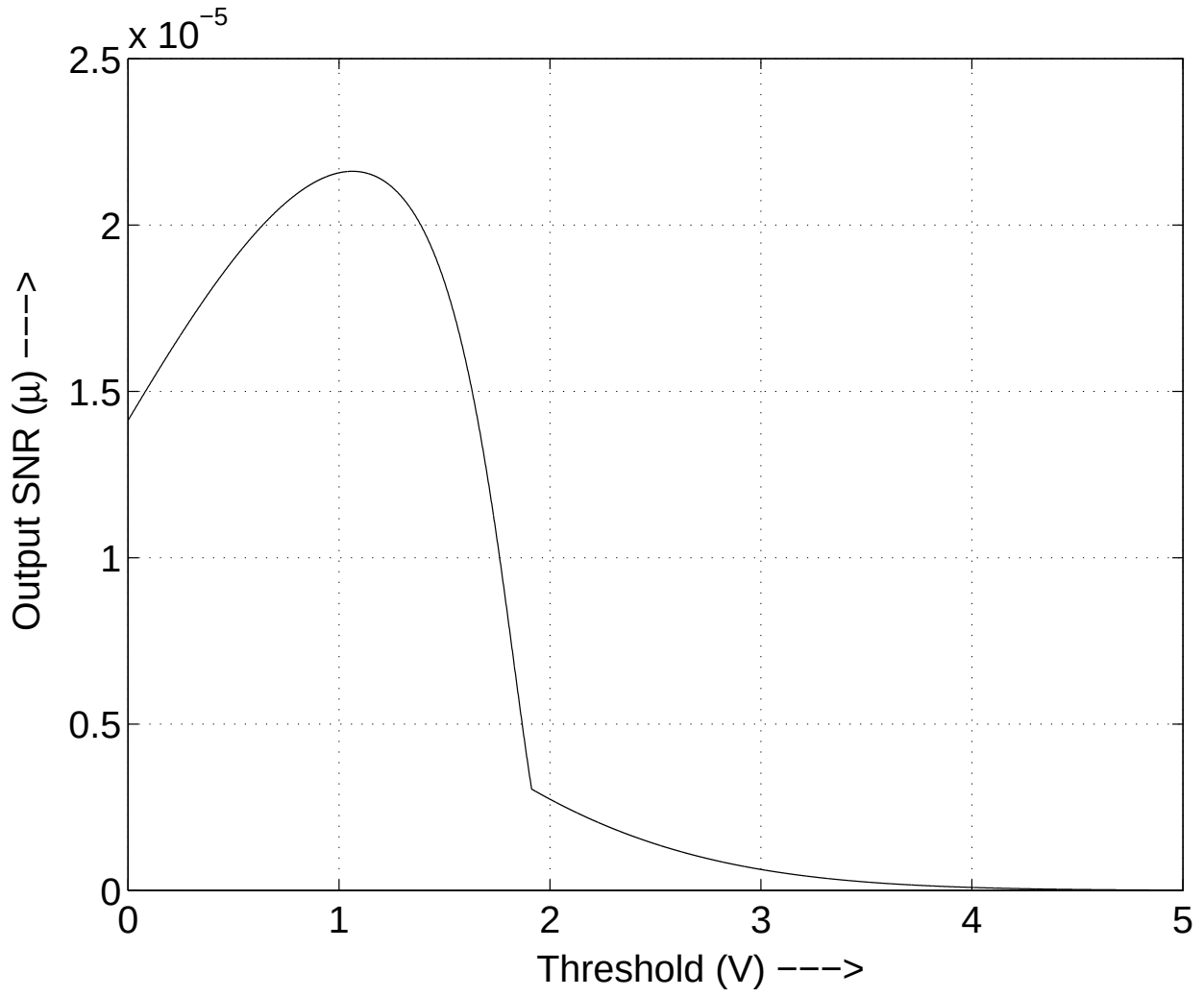

Fig. 5. Plot of equation 19 for output SNR $\mu$ vs threshold $\gamma$. The maximum SNR occurs around $\gamma = 1.064$.

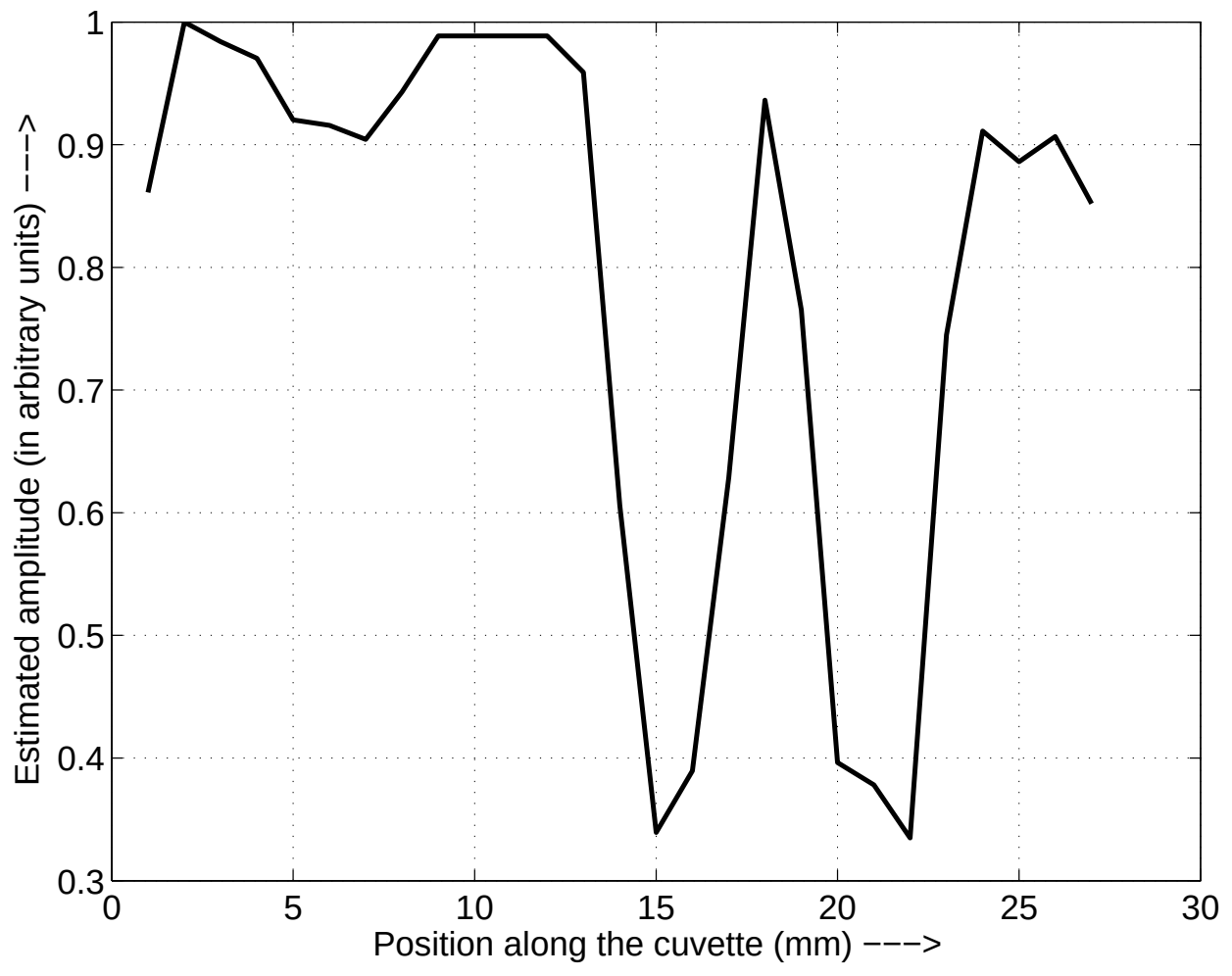

Fig. 6. The plot of the estimated amplitude vs position of the PMT along the cuvette in the case of lock-in detection. The turbidity level is given by $L/l^* = 2.14$. The presence of the glass rod is clearly visible.

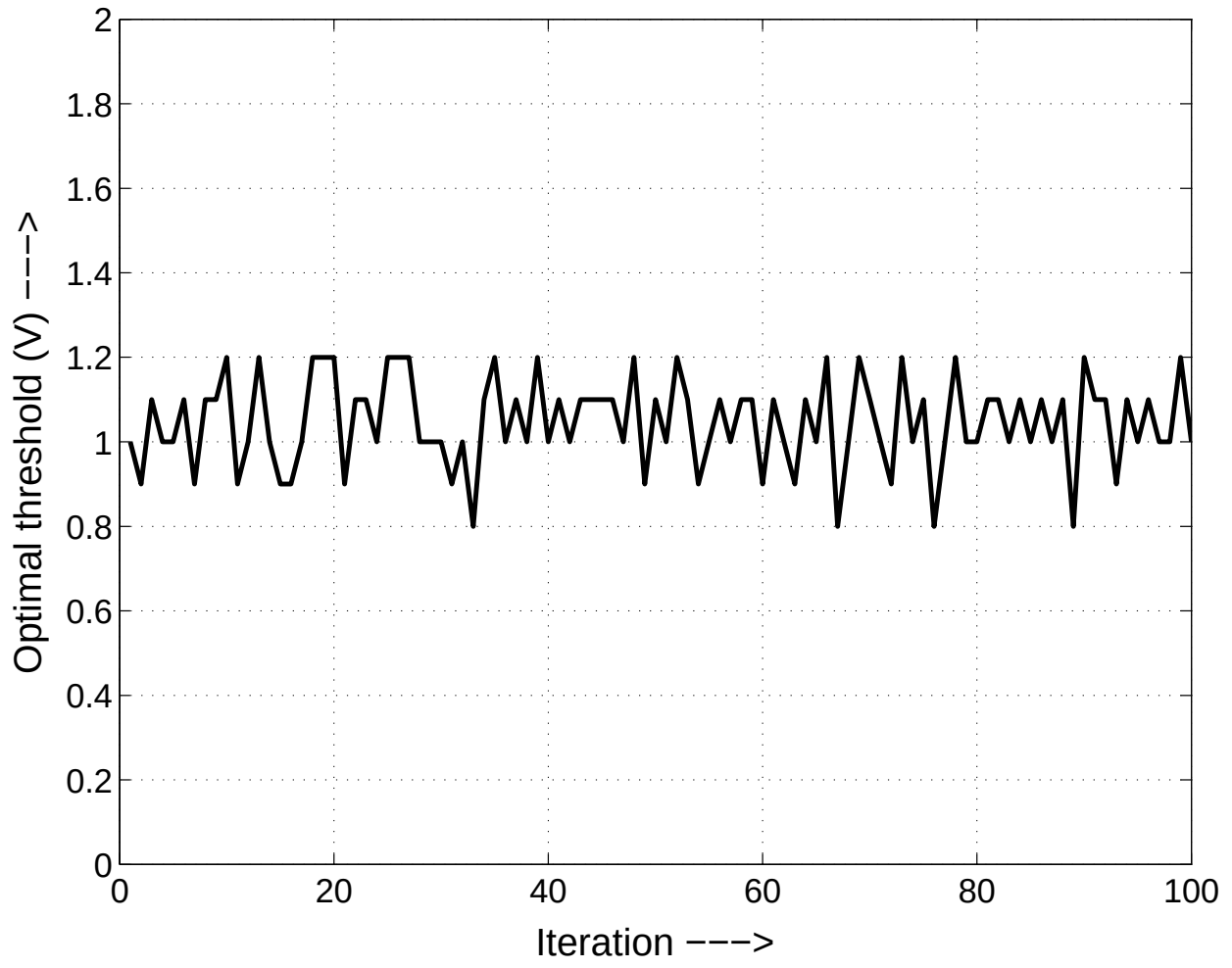

Fig. 7. Plot of the optimal threshold for the quantizer obtained for one hundred samples of the simulated data. The SNR in the data is kept at -23dB and the noise variance is kept at unity. The average value of the optimal threshold obtained is 1.046 which can be compared to the value obtained from eqn. (20) which is 1.064.

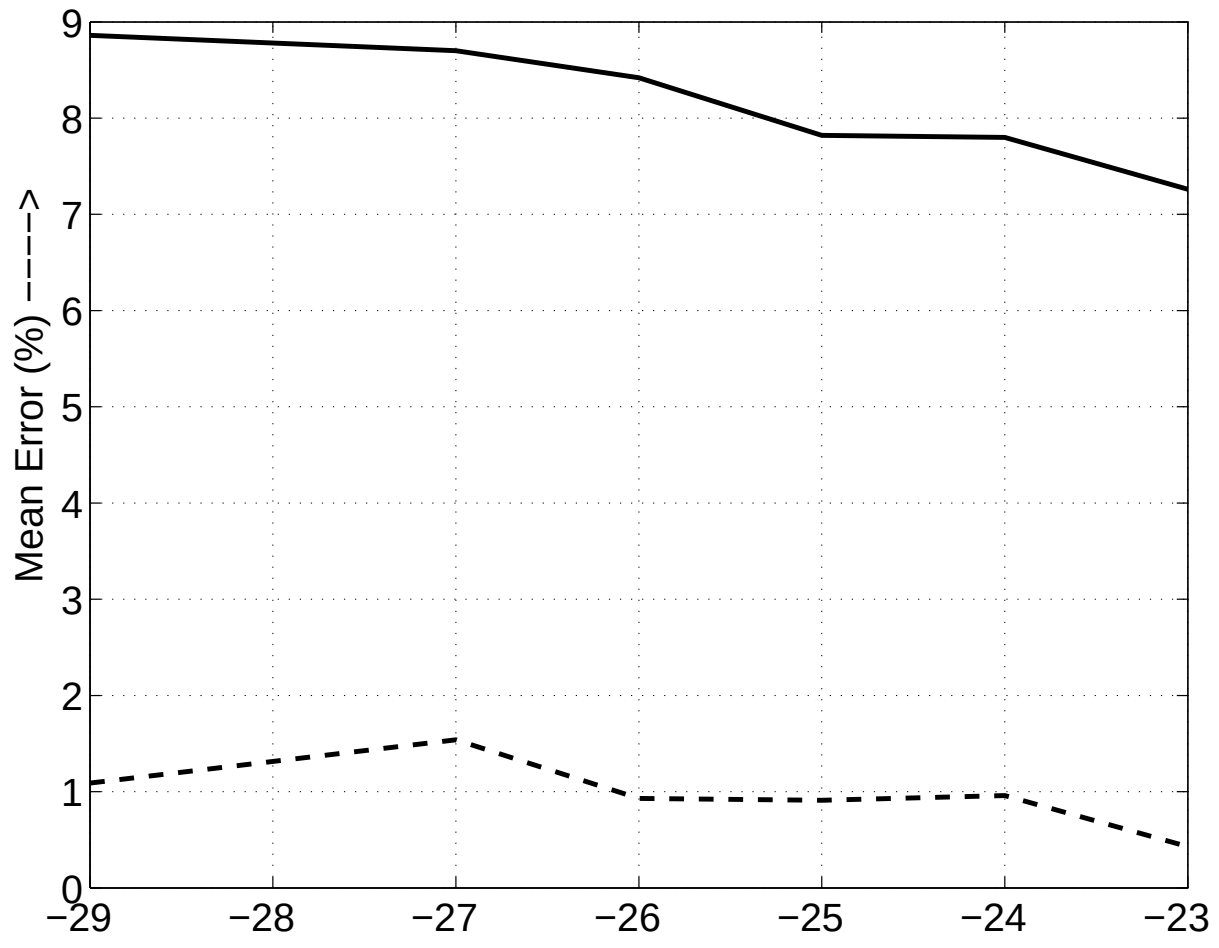

Fig. 8. The plot of the percentage error in the estimated amplitude vs SNR in the input from the 3 level quantizer employing the crossover probability method. The dotted line represents the case when the frequency of the sinusoid is known exactly. The solid line is for the case when frequency is known only within an error of 0.05%.

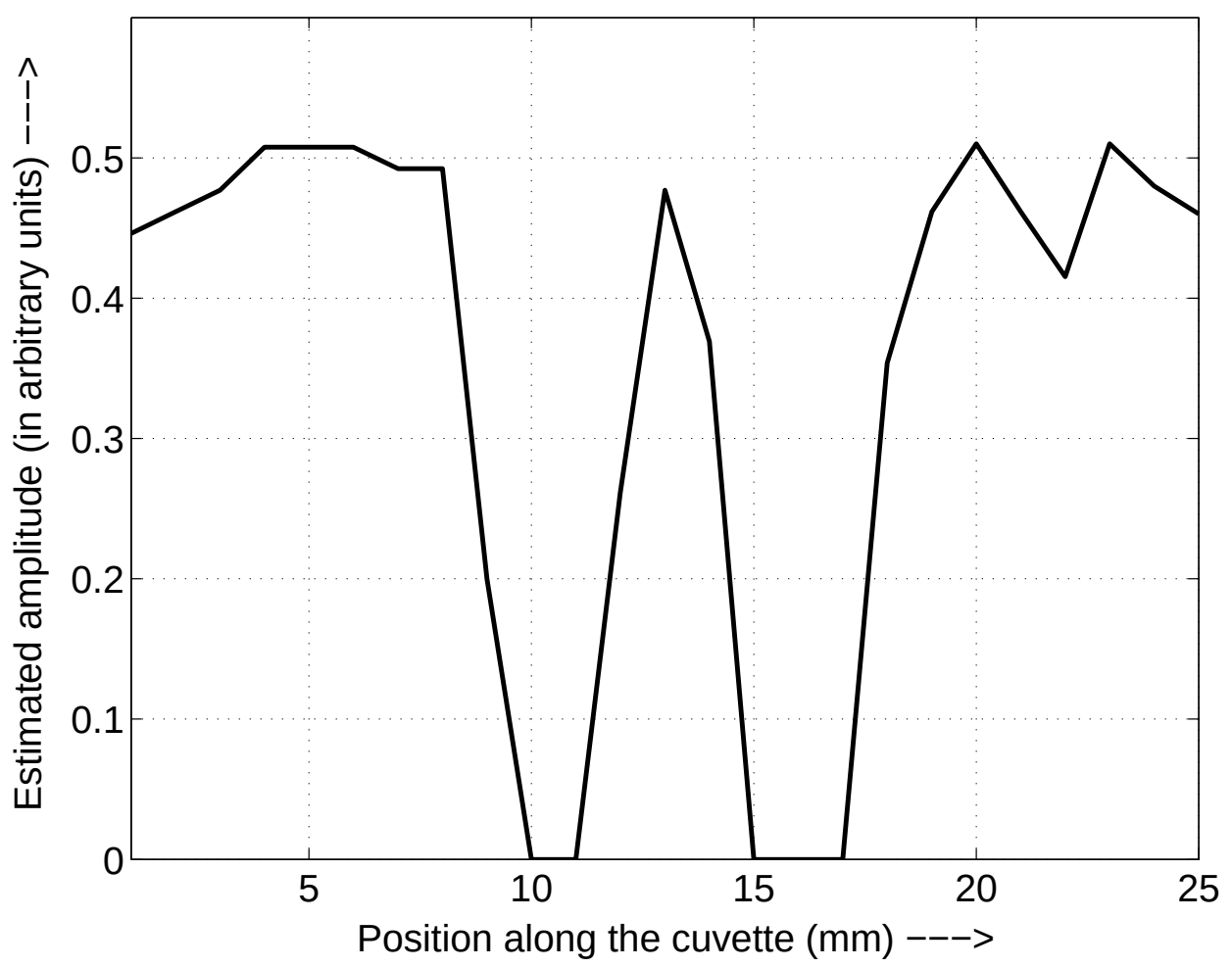

Fig. 9. The plot of the estimated amplitude vs position of the PMT along the cuvette in the case of amplitude estimated from the 3 level quantizer employing cross-over probability method. The turbidity level is given by $L/l^* = 2.14$. The presence of the glass rod is clearly visible.

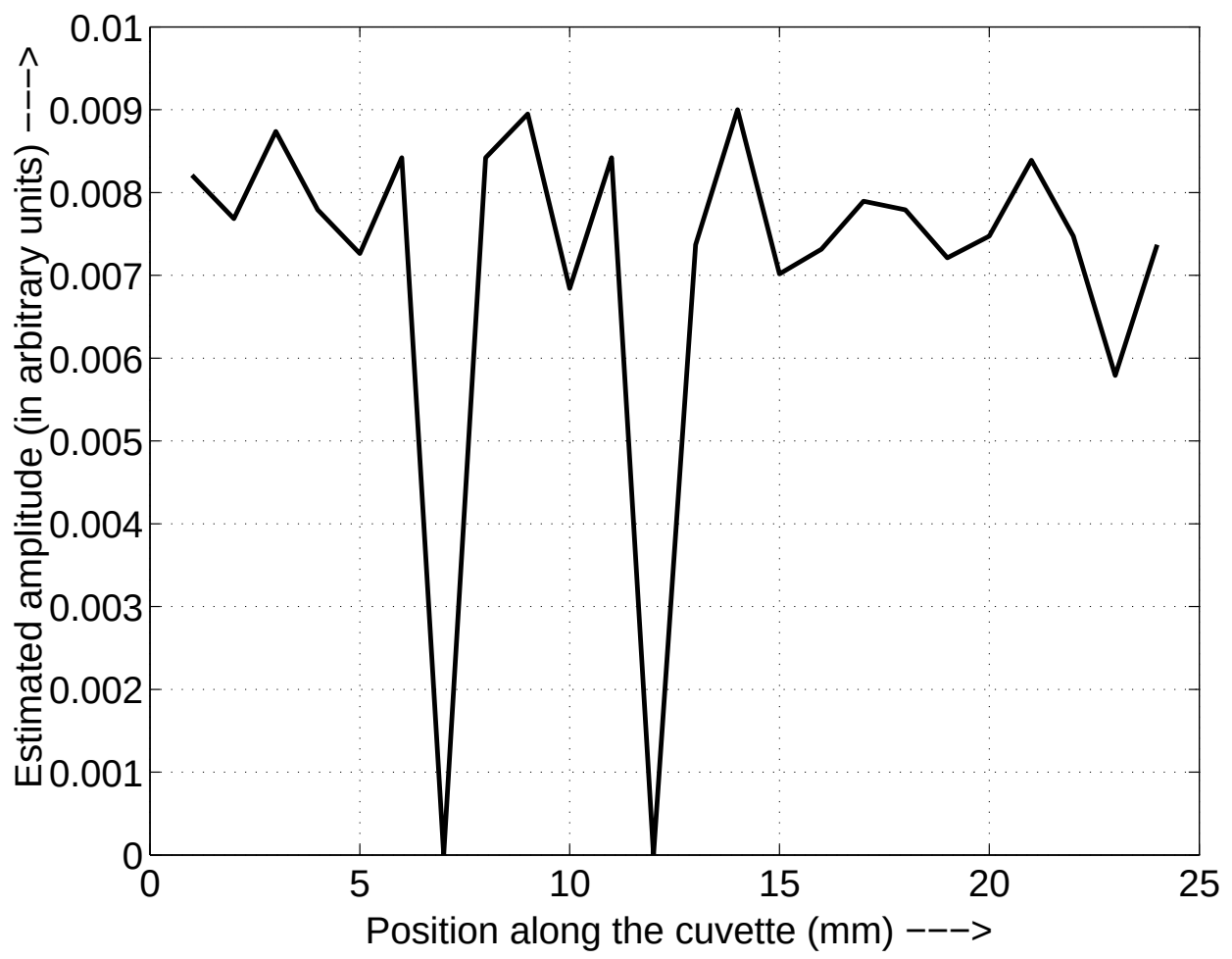

Fig. 10. The same as Figure 9 except $L/l^* = 5.05$. The presence of the glass rod is still clearly visible.

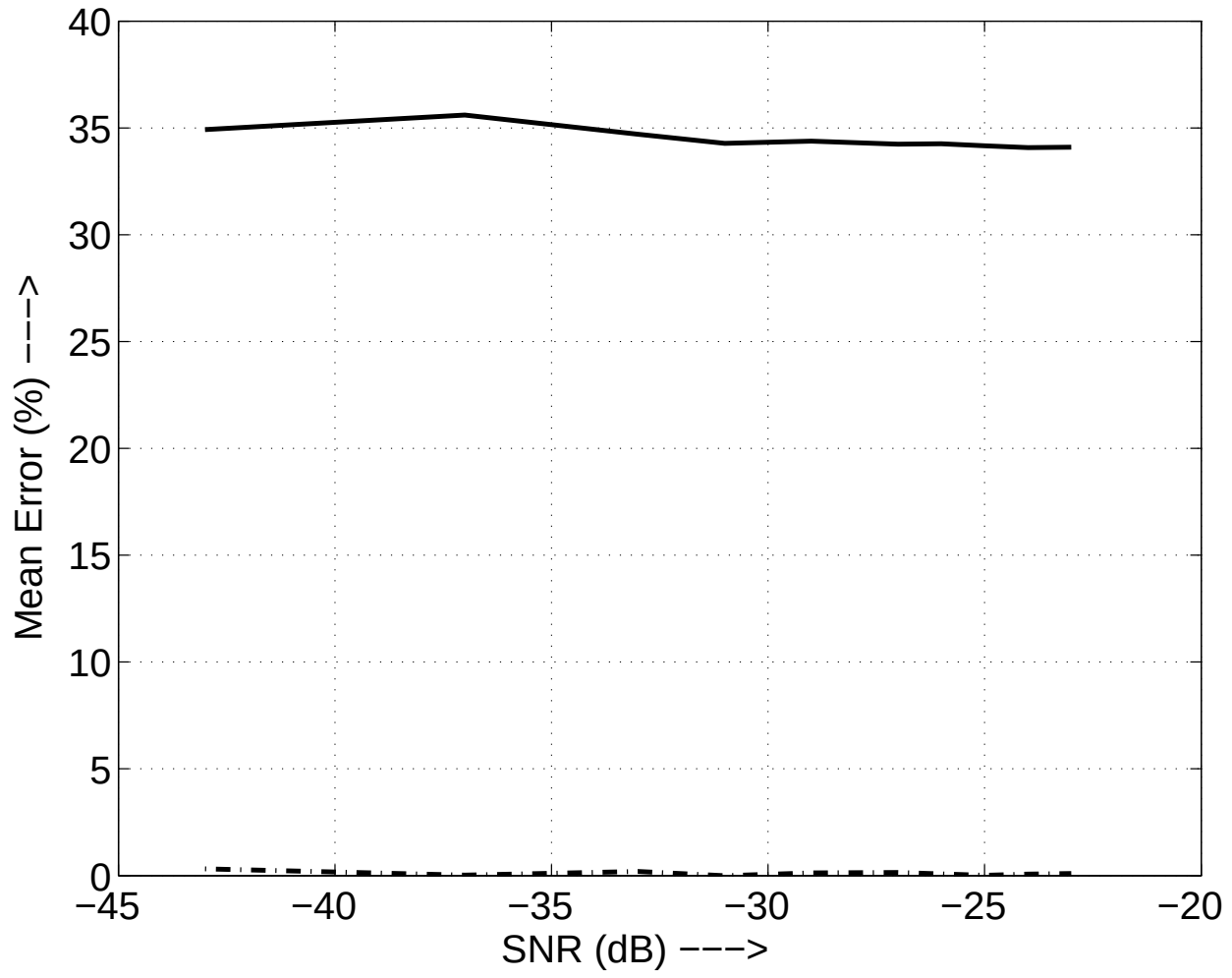

Fig. 11. The plot of the percentage error in the estimated amplitude vs SNR in the input for the method employing the MLE for quantized data. The dotted line represents the case when the frequency of the sinusoid is known exactly. The solid line is for the case when frequency is known only within an error of 0.05%.

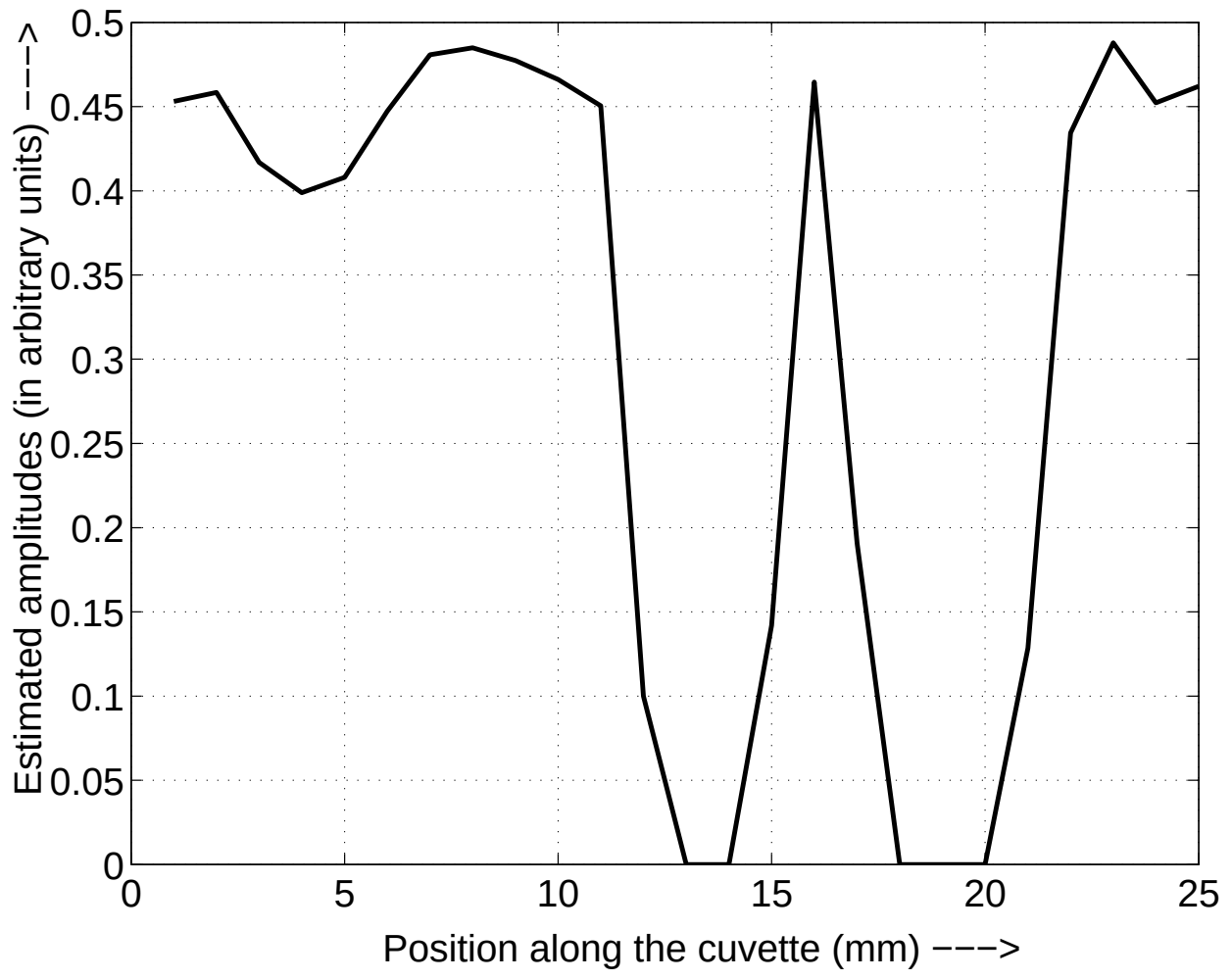

Fig. 12. The plot of the estimated amplitude vs position of the PMT along the cuvette in the case of amplitude estimated from the MLE for quantized data. The turbidity level is given by $L/l^* = 2.14$. The presence of the glass rod is clearly visible.

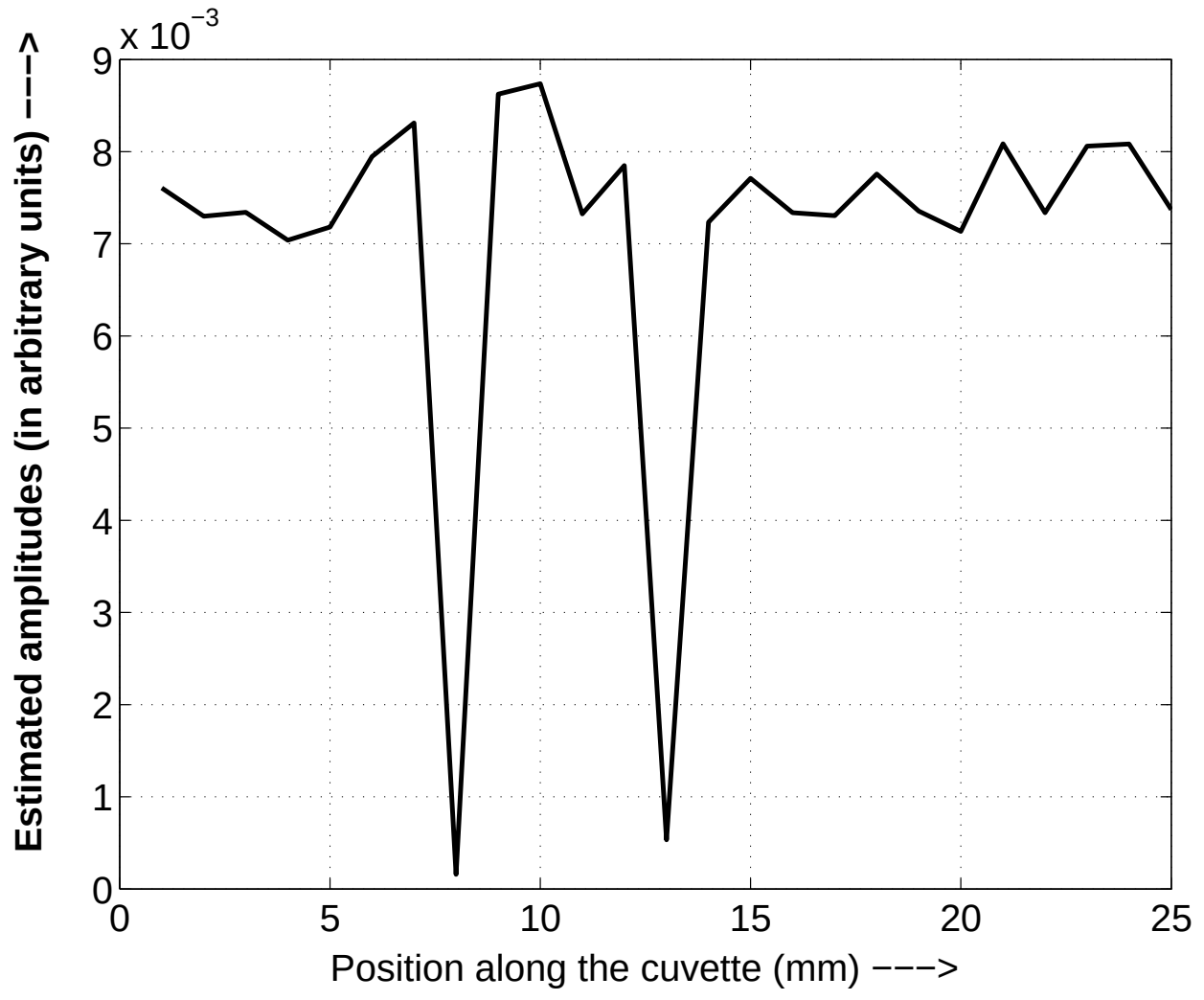

Fig. 13. The same as Figure 12 except $L/l^* = 5.05$. The presence of the glass rod is still clearly visible.

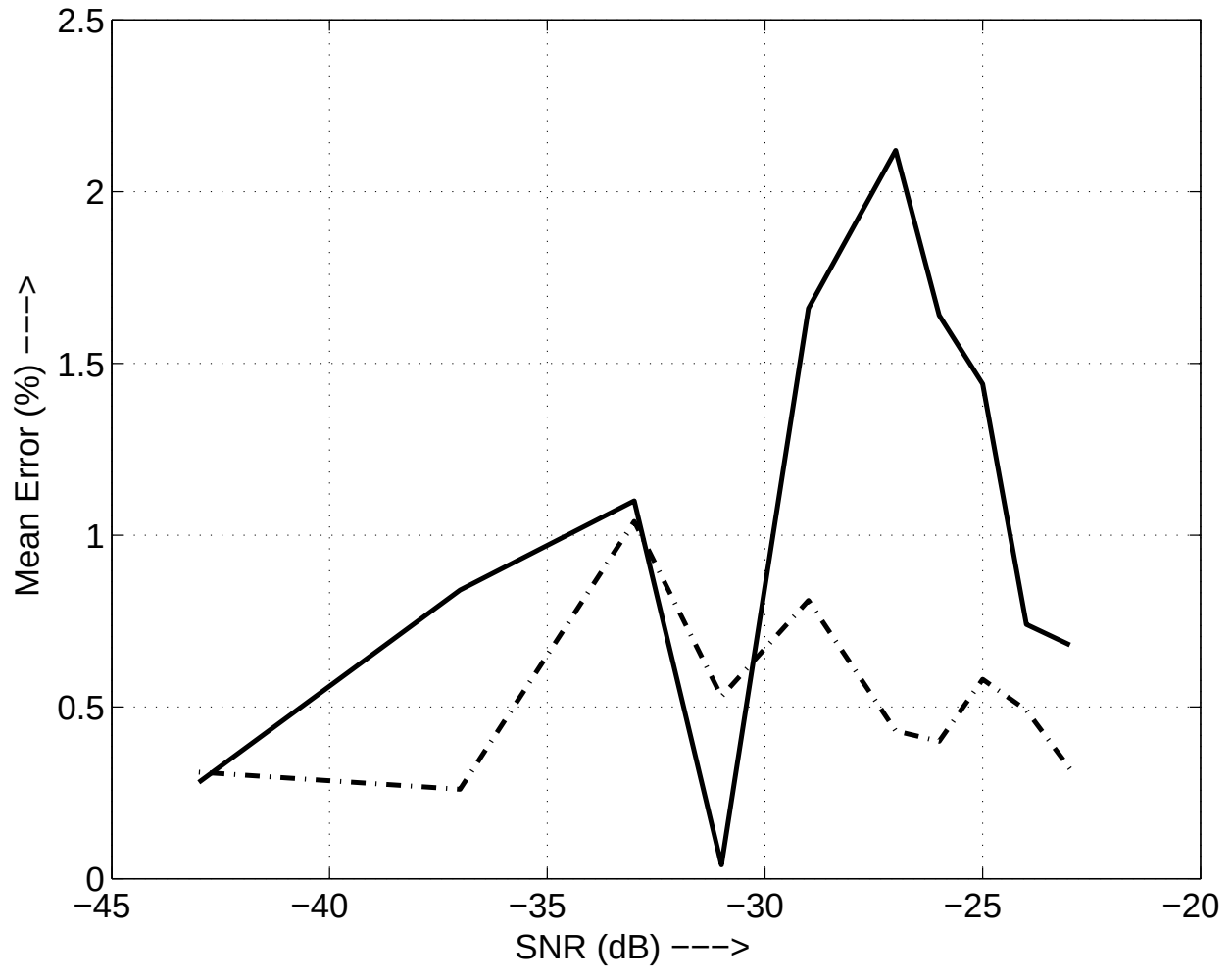

Fig. 14. The plot of the percentage error in the estimated amplitude vs SNR in the input employing the expected power of the quantized output. The dotted line represents the case when the frequency of the sinusoid is known exactly. The solid line is for the case when frequency is known only within an error of 0.05%.

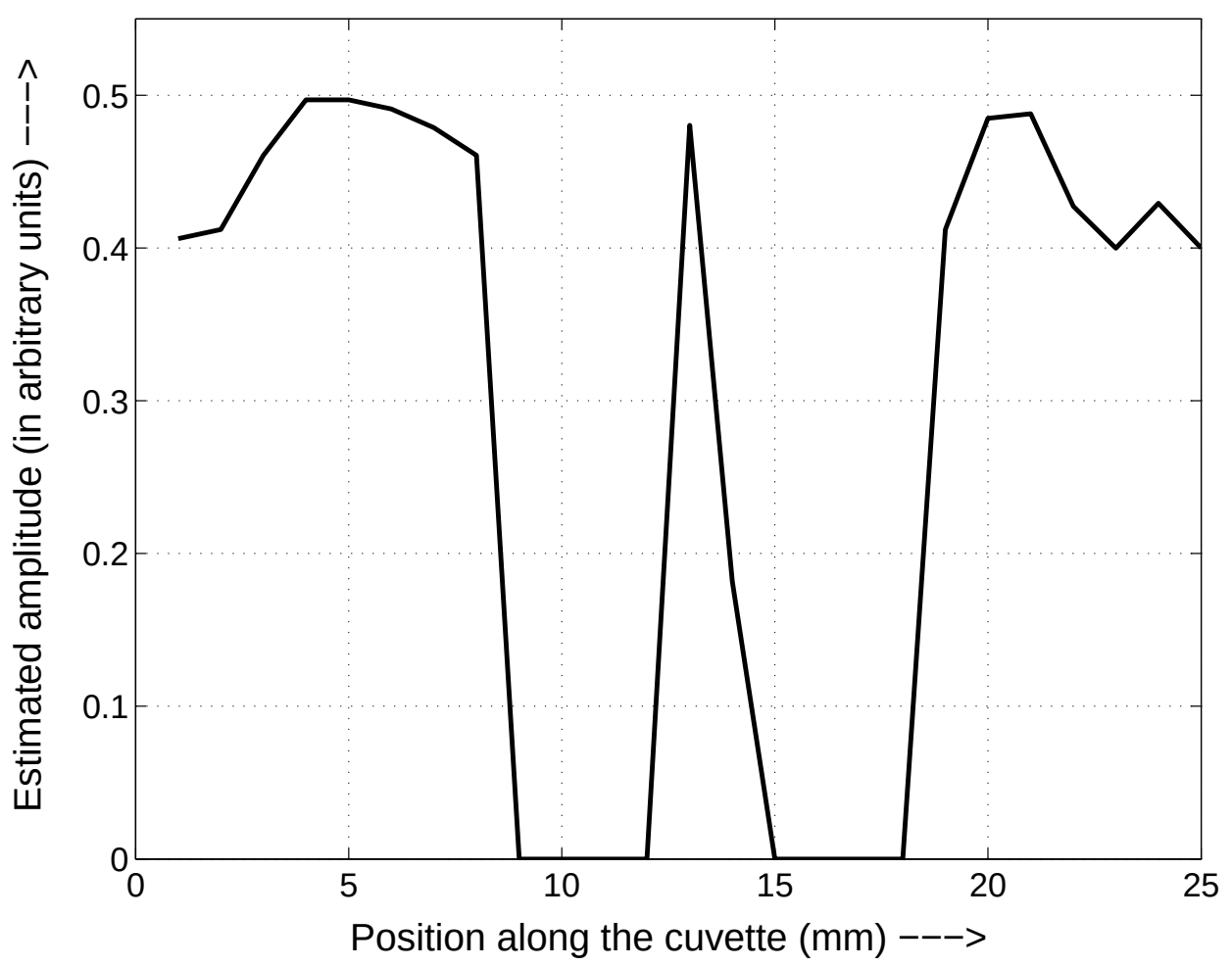

Fig. 15. The plot of the estimated amplitude vs position of the PMT along the cuvette in the case of amplitude estimated from the 3 level quantizer employing the expected power of the quantized output method . The turbidity level is given by $L/l^{*} = 2.14$. The presence of the glass rod is clearly visible.

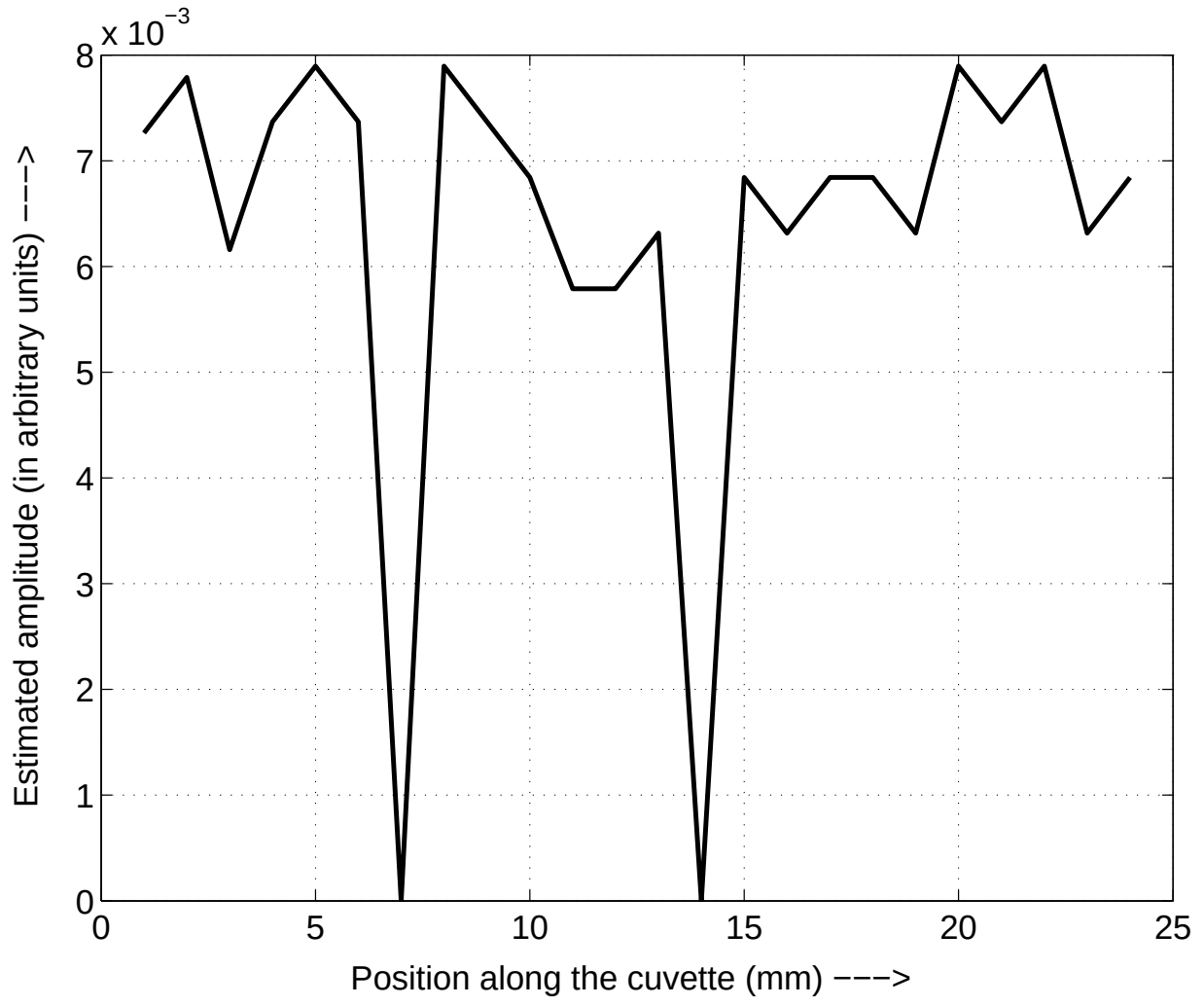

Fig. 16. The same as Figure 15 except $L/l^{*} = 5.05$. The presence of the glass rod is still clearly visible.